\documentclass[aps,prl,twocolumn,groupedaddress]{revtex4}
\usepackage[dvips]{graphicx}
\begin{document}
\title{
%
%
\[ \vspace{-2cm} \]
\noindent\hfill\hbox{\rm Alberta Thy 15-02} \vskip 5pt
%
%
Towards a standard jet definition}
\author{D.\ Yu.\ Grigoriev$^{1,3}$, E.\ Jankowski$^2$,
and F.\ V.\ Tkachov$^3$} 
\affiliation{
$^1$
Mathematical Physics,
Natl.\ Univ.\ of Ireland Maynooth,
Maynooth, Co.\ Kildare, Ireland\\
$^2$
Department of Physics,
University of Alberta,
Edmonton, AB, T6G 2J1, Canada\\
$^3$
Institute for Nuclear Research of RAS,
Moscow 117312, Russia}
\date{\today}
\begin{abstract}
In a simulated measurement of the $W$-boson mass,
evaluation of Fisher's information shows the optimal jet definition
\cite{r1} to yield the same precision as the $k_\mathrm{T}$ algorithm 
while being much faster at large multiplicities.
\end{abstract}
\maketitle

\textbf{1.}\
Association of hadronic jets observed in high energy physics experiments
with quarks and gluons in the underlying collisions of quanta \cite{r2}
provides an experimental handle on fundamental interactions via the so-called
jet finding algorithms that find 
a configuration of jets $\mathbf{Q}$, represented by the $N_\mathrm{jets}$
4-momenta $p_j$, for a given event $\mathbf{P}$,
represented by the $N_\mathrm{part}$ light-like 4-momenta $p_a$:
%
\begin{equation}\label{eq-1}
\mathbf{P}=
\left\{p_a\right\}
\stackrel{\mathrm{jet\;algorithm}}
{-\!\!\!-\!\!\!-\!\!\!-\!\!\!-\!\!\!-\!\!\!-\!\!\!-\!\!\!-\!\!\!
\longrightarrow}
\mathbf{Q}=
\left\{p_j\right\}.
\end{equation}
%
Unless jets are energetic and well separated, jet definition
involves ambiguities that were seen to be a major, even dominant source
of errors in the planned experiments \cite{r3}.

A well-known requirement on possible jet finding algorithms
is the infrared safety \cite{r4} or insensitivity of $\mathbf{Q}$ to 
collinear fragmentations of particles in $\mathbf{P}$.
It is clarified by a theorem of ref.\ \cite{r5} expressing
fragmentation-invariant observables in terms of the
energy-momentum tensor defined by space-time symmetries uniquely,
so that such observables can be equivalently represented
in terms of either hadron, or quark and gluon fields. 
However, the requirement leaves much freedom for the mapping (\ref{eq-1}),
and many jet algorithms emerged over time. 

\textbf{2.}\
Ref.\ \cite{r4} introduced so-called cone algorithms that define a jet
as all particles in a cone of a fixed radius \cite{r6}.
Cone axes are usually found iteratively to be directed along jets'
3-momenta, and cone overlaps are treated with ad hoc prescriptions.
The fixed shape of cones enhances the stability of cone algorithms
and facilitates studies of detector corrections, 
but decreases the jet resolution power.

Ref.\ \cite{r7} introduced a definition based on the shape observable thrust
\cite{r8}, as theoretical studies are easier with such observables.
Here one minimizes the sum
%
\begin{equation}\label{eq-2}
\textstyle\sum_j \, \left( {1 - T_j } \right),
\end{equation}
%
where $T_j$  is the thrust for the $j$-th jet. (Similar measures
were considered e.g.\ in \cite{r33} and a special case of jet search
based on an optimization of a shape observable was also employed
e.g.\ in \cite{r34}.) 
However, the required minimization was deemed unfeasible \cite{r9}.

Successive recombination algorithms emerged with a motivation to invert
hadronization \cite{r10}.
Here one starts with a list of particles, computes a ``distance'' $d_{ab}$
for each pair of particles, and replaces the pair with the smallest $d_{ab}$
by a single pseudo-particle with $p_{ab}=p_a+p_b$.
One repeats this until e.g.\ all $d_{ab}$ exceed a given threshold
$y_\mathrm{cut}$ or only a given number of (pseudo-) particles remain
in the list. Possible $d_{ab}$ are given by
%
\begin{equation}\label{eq-3}
d^2_{ab}=E_a E_b \left(E_a+E_b\right)^{-n} \left(1-\cos\theta_{ab}\right),
\end{equation}
%
where $E_a$ and $E_b$ are the particles' energy fractions,
$\theta_{ab}$ is the angle between them, and $y_\mathrm{cut}$
is the so-called jet resolution parameter.
$n=0$ and $n=2$ correspond to the JADE \cite{r11}
and Geneva \cite{r12} criteria.
It was also argued that the dynamics of the $2 \to 1$ amplitude in 
QCD is matched best by the so-called $k_\mathrm{T}$ measure \cite{r13}:
%
\begin{equation}\label{eq-3a}
d^2_{ab}=\min\left(E^2_a, E^2_b\right)\left(1-\cos\theta_{ab}\right).
\end{equation}
%

Such algorithms find jets of irregular shapes.
Ref.\ \cite{r14} replaced $2 \to 1$ recombinations with a global $n \to m$ one
(but still based on pairwise distances $d_{ab}$),
yielding more regular jets but this is more expensive computationally.

The multitude of available jet algorithms
--- often differing in obscure details --- caused their comparative studies
(e.g.\ refs.\ \cite{r6}, \cite{r12}, \cite{r10}, \cite{r15}).
The subject's importance has been growing along with the drive
towards higher precision in the jet physics \cite{r3}, \cite{r15}.

\textbf{3.}\
Ref.\ \cite{r16} reinterpreted the physically significant ambiguities
of jet algorithms due to algorithmic variations
as instabilities which a correct measurement procedure must be free of.
The resulting theory \cite{r1}, \cite{r17} provided a context to derive
an optimal jet definition from explicit physical motivations. 
The principal  points of the theory are as follows:

(i) Calorimetric measurements with hadronic final states $\mathbf{P}$
must rely on observables $f\left(\mathbf{P}\right)$
that possess a special "calorimetric", or $C$-continuity which is
a non-perturbative generalization of the familiar IR safety
(see \cite{r17} for details) and which guarantees a stability
of $f\left(\mathbf{P}\right)$ against distortions of $\mathbf{P}$
such as caused by detectors. Ref.\ \cite{r17} pointed out $C$-continuous
analogues for a variety of observables usually studied via
intermediacy of jet algorithms.
The fundamental role of such observables is highlighted by two facts:
(1) An observable inspired by \cite{r17} played an important role in
the selection of top quark events in the fully hadronic channel at D0
\cite{r19}, \cite{r20}.
(2) The Jet Energy Flow project \cite{r21} provides numerical evidence that
$C$-continuous observables may indeed help to go beyond the intrinsic
limitations of conventional procedure based on jet algorithms in the quest
for the 1\% precision level in the physics of jets.

(ii) The proposition that the observed event $\mathbf{P}$ inherits
information (as measured by calorimetric detectors) from the underlying
quark-and-gluon event $\mathbf{q}$ is expressed as 
%
\begin{equation}\label{eq-4}
f\left(\mathbf{q}\right)
\approx
f\left(\mathbf{P}\right)
\mathrm{\quad for \;any \;}C\mathrm{\!-\!continuous}\;f.
\end{equation}
%

(iii) For each parameter $M$ on which the probability distribution
$\pi_M\left(\mathbf{P}\right)$ of the observed events $\mathbf{P}$ may depend,
there exists an optimal observable
$f_\mathrm{opt} \left(\mathbf{P}\right) =
\partial_M \ln \pi_M \left(\mathbf{P}\right)$
for the best possible measurement of $M$ \cite{r18}. 
This is a reinterpretation of the Rao-Cramer
inequality and the maximal likelihood method of mathematical statistics
in terms of the method of moments. In the context of multi-hadron final
states as "seen" by calorimetric detectors,
such an observable is automatically $C$-continuous.

(iv) If the dynamics of hadronization is such that eq.\ (\ref{eq-4}) holds,
then good approximations for $f_\mathrm{opt}$ could exist among functions
that depend only on $\mathbf{Q}$ which is a parameterization
of $\mathbf{P}$ in terms of a few pseudo-particles
(jets), found from a condition modeled after eq.\ (\ref{eq-4}):
%
\begin{equation}\label{eq-5}
f\left(\mathbf{Q}\right) \approx f\left(\mathbf{P}\right)
\mathrm{\quad for \;any \;}C\mathrm{\!-\!continuous}\;f.
\end{equation}
%
This simply translates the meaning of jet finding as an inversion of
hadronization into the language of $C$-continuous observables.

(v) $C$-continuous observables can be approximated by sums of products
of simplest such observables that are linear in particles' energies: 
%
\begin{equation}\label{eq-6}
\textstyle f\left(\mathbf{P}\right) =
\sum_a E_a f\left(\mathbf{\hat{p}}_a\right).
\end{equation}
%
(The relevant theorems can be found in refs.\ \cite{r1} and \cite{r17}.)

(vi) So it is sufficient to explore the criterion (\ref{eq-5}) with only
$f$'s of the form (\ref{eq-6}). Then one can perform a Taylor expansion
in angular variables and obtain a factorized bound of the form 
%
\begin{equation}\label{eq-7}
\left| f\left(\mathbf{P}\right) - f\left(\mathbf{Q}\right) \right|
\le C_{f,R}
\times \Omega_R \left[\mathbf{P},\mathbf{Q}\right],
\end{equation}
%
where all the dependence on $f$ is localized within $C_{f,R}$ 
(so the bound remains valid for any $C$-continuous $f$) and
%
\begin{equation}\label{eq-8}
\Omega_R \left[\mathbf{P},\mathbf{Q}\right]=
R^{-2} \mathrm{Y} \left[\mathbf{P},\mathbf{Q} \right]
+\mathrm{E}_\mathrm{soft} \left[\mathbf{P},\mathbf{Q} \right],
\end{equation}
%
where
$\mathrm{Y}\left[\mathbf{P},\mathbf{Q}\right]=
2 \sum_j p_j \tilde{q}_j$,
$\mathrm{E}_\mathrm{soft} \left[\mathbf{P},\mathbf{Q}\right] =
\sum_a\bar{z}_a E_a$, and $R>0$ is a free parameter (see ref.\ \cite{r1}
for a discussion).
$p_j$ are jets' physical 4-momenta expressed as
$p_j  = \sum_a z_{aj} p_a$,
where the so-called recombination matrix $z_{aj}$ is such that
$0 \le z_{aj}  \le 1$ and
$\bar{z}_a  = 1 - \sum_j z_{aj}  \ge 0$
for any $a$, i.e.\ a part of the particle's energy is allowed to not 
participate in any jet. 
$\tilde{q}_j$ are light-like 4-vectors related to $p_j$ and given by
$\tilde{q}_j=\left(1, \mathbf{p}_j/ \left|\mathbf{p}_j\right|\right)$
for lepton collisions ($\tilde{q}_j$ can be definied differently
for hadron collisions; see ref. \cite{r1} for details).
The recombination matrix $z_{aj}$ occurs naturally in the construction
of the bound (\ref{eq-7}) and is the fundamental unknown in this scheme.
$\mathrm{Y}$ in (\ref{eq-8}) differs from (\ref{eq-2})
in that the jet's physical momentum is used in place of the thrust axis.
$E_\mathrm{soft}$ is the event's energy fraction that does not take part
in jet formation. 
(vii) Since the collection of values of all $f$ on a given event
$\mathbf{P}$ is naturally interpreted as the event's physical information
content, the bound (\ref{eq-7}) means that the distortion of such content
in the transition from $\mathbf{P}$ to $\mathbf{Q}$ can be controlled
by a single function; so the loss of physical information in the transition
is minimized if $\mathbf{Q}$ corresponds to the global minimum of $\Omega_R$.
The Optimal Jet Definition amounts to finding $z_{aj}$ which minimizes
$\Omega_R$, depending on specific application, either with a given number
of jets or with a minimum number of jets while satisfying the restriction 
$\Omega_R \left[\mathbf{P},\mathbf{Q}\right]<\omega_\mathrm{cut}$ 
with some parameter $\omega_\mathrm{cut}>0$ which is similar to the jet
resolution $y_\mathrm{cut}$ of recombination algorithms.

\textbf{4.}\
OJD combines attractive features of the different algorithms
reviewed above and is free of their defects (see ref.\ \cite{r1}
for more details): 
(i) OJD is based on a shape observable. 
(ii) It finds jets of rather regular shapes with angular radii bounded by $R$.
(iii) it resolves jet overlaps dynamically, depending on the global structure
of the event's energy flow.
(iv) $\omega_\mathrm{cut}$ bounds the soft energy in the physically preferred
totally inclusive fashion (cf.\ ref.\ \cite{r4}.
(v) OJD is purely analytical, allowing its algorithmic implementations
to differ beyond programmatic code optimizations and to be customized
for specific applications.
(vi) OJD is embedded in a systematic theory with new options
for constructing improved data processing procedures that go beyond
the conventional approach. 

\textbf{5.}\
Despite the huge dimension of the domain in which to search the global
minimum,
$N_\mathrm{part} \times N_\mathrm{jets}=
O\left(100\mathrm{-}1000\right),$
OJD lends itself to efficient algorithmic
implementations (the Optimal Jet Finder library \cite{r23}).

OJF was first developed in the programming language Component Pascal
\cite{r25}, featuring a unique combination of safety and efficiency.
This was very useful for the experimentation needed to find a satisfactory
algorithm.
Only after that the final port to FORTRAN was performed. 
A subsequent testing \cite{r26} and a substantially
independent verification \cite{r27} revealed no defects 
of significance, indicating a high reliability of the resulting code
\cite{r24}.

The OJF library can be used to obtain OJD implementations adapted for
specific applications (see below).

\textbf{6.}\
A number of successive recombination algorithms were compared in
ref.\ \cite{r9} in a series of tests none of which, however, was conclusive.
The JADE algorithm proved to be the least satisfactory, the Geneva algorithm
behaved somewhat erratically, and a group of algorithms
(including $k_\mathrm{T}$ and Luclus) exhibited a balanced behavior in various
tests, typically populating the spread between the JADE and Geneva algorithms.
Note that the successive recombination scheme is recovered within OJD as
a heuristic minimum-search trick with $n=1$ in eq.\ \ref{eq-3} \cite{r17},
which is the geometric mean of the JADE and Geneva criteria. Then OJD
should roughly fall into the same group as the $k_\mathrm{T}$
and Luclus algorithms. 
A conclusive physically meaningful comparison can be performed
in the context of the method of optimal observables.
We explain the procedure using a simple example modeled after
the measurements of the $W$-boson mass $M$ at LEP2 \cite{r28}.
The details will be published separately \cite{r29}.

The process $e^+e^- \to W^+W^- \to \mathrm{hadrons}$ at CM energy
of 180 GeV was simulated using PYTHIA 6.2 \cite{r30}.
Each event was resolved into 4 jets. These can be combined into two pairs
(supposedly resulting from decays of the $W$'s) in three different ways;
we chose the combination with the smallest difference in invariant masses
between the two pairs and calculate the average $m$ of the two masses.
This mapped events to the $m$ axis. We used $9\cdot10^6$ events
to generate the probability distribution $\pi _M (m)$
and to construct a numerical approximation to the optimal observable
$f_\mathrm{opt}\left(m\right) = \partial_M \ln\pi_M\left(m\right)$. 
Using this as a generalized moment with a sample of
$N_\mathrm{exp}$ experimental
events would yield an estimate for $M$ with the theoretically 
smallest error estimated as
$\delta M_\mathrm{exp} \cong
\left(N_\mathrm{exp}\left\langle f_\mathrm{opt}^2\right\rangle \right)^{-1/2}$,
where
$\left\langle f_\mathrm{opt}^2 \right\rangle$ is sometimes 
identified with Fisher's information.
$\delta M_\mathrm{exp}$ immediately reflects suitability of the jet
algorithm used.

We thus compared OJD with the $k_\mathrm{T}$ and JADE definitions.
We used the KTCLUS implementation of the $k_\mathrm{T}$ algorithm \cite{r31}
and modified the recombination criterion to obtain the JADE algorithm.
All events were forced to 4 jets, so the parameters $y_\mathrm{cut}$
and $\omega_\mathrm{cut}$ played no role.

\begin{figure}[htb]
\begin{center}
\includegraphics[width=8cm]{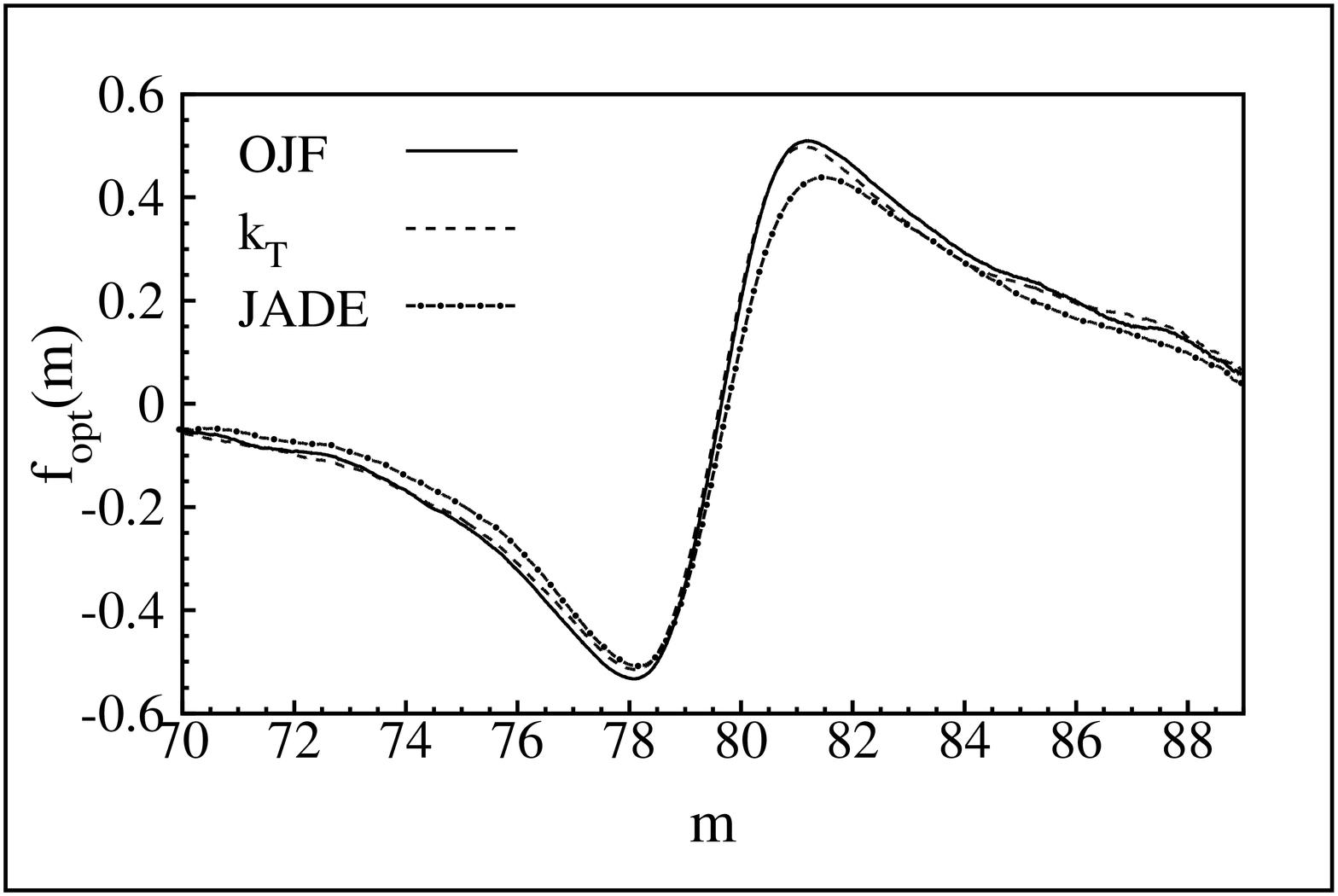}
\caption{Optimal observable $f_\mathrm{opt}\left(m\right)$
for OJF, $k_\mathrm{T}$ and JADE.}
\label{fig}
\end{center}
\end{figure}

For OJD, we chose $R\!=\!2$ and, for benchmarking purposes, first employed
a primitive variant of OJF-based algorithm with a fixed
$n_\mathrm{tries}$ for all events, where $n_\mathrm{tries}$ is
the number of independent attempts to descend into a global minimum
from a random initial configuration. The probability to miss
the global minimum vanishes for larger $n_\mathrm{tries}$;
we chose $n_\mathrm{tries}=10$. The obtained $f_\mathrm{opt}\left(m\right)$
for the three jet algorithms are shown in fig.\ \ref{fig}.

For $N_\mathrm{exp}=1000$
(which roughly corresponds to the $W$-mass measurements at LEP2) we found
the following:
\begin{center}
\begin{tabular}{|ccc|}
\hline
ALGORITHM &\phantom{100}& $\delta M_\mathrm{exp}\pm3\;\mathrm{MeV}$\\
\hline
OJD/OJF &\phantom{100}& 106\\
$k_\mathrm{T}$ &\phantom{100}& 105\\
JADE &\phantom{100}& 118\\
\hline
\end{tabular}\\
\end{center}
The error of 3 MeV is mostly due to numerical
differentiation in $M$.

Note that there are options to improve the measurement procedure that are
specific to OJD, e.g.\ weighting events according to the values of
$\Omega_R$. We have not explored them, as it is sufficient for the purposes
of this Letter to establish that OJD is at least no worse than
the $k_\mathrm{T}$ algorithm for this measurement.  

\textbf{7.}\
An important aspect is the speed of jet algorithms at large
$N_\mathrm{part}$. This is critical e.g.\ in the preclustering 
for reducing the number of clusters in each event as seen e.g.\
by the D0 detector at FNAL to about 200: otherwise it is 
not possible to analyze data with $k_\mathrm{T}$ as its processing time
per event is $O\left(N_\mathrm{part}^3\right)$ \cite{r32}.
A concern then is how the preclustering affects the final results
as it has to be done using a method unrelated to the $k_\mathrm{T}$ 
algorithm, and a non-programmatic modification of the latter 
must be treated as a new jet definition (cf.\ examples in ref.\ \cite{r9}).

Speed of the algorithms as different as OJF and \mbox{KTCLUS}
(coded in the same dialect of FORTRAN) may depend on the 
computing installation.
With this in view, we report times per event in units of $10^{-2}$ sec 
as measured on our hardware with our sample of events.

$N_\mathrm{part}$ varied from 50 to 170 in our sample, with the mean value
of 83. The processing time per event is described rather well
by the following formulae:
%
\begin{equation}\label{eq-10}
\begin{array}{ll}
1.2 \times 10^{-6} \times N_\mathrm{part}^3
&\;\mathrm{for\;KTCLUS},\\ 
1.0 \times 10^{-2} \times N_\mathrm{part}  \times n_\mathrm{tries}
&\;\mathrm{for\;OJF}.
\end{array} 
\end{equation}
%
This behavior was verified for $N_\mathrm{part}$ up to 1700 by splitting
each particle into 10 collinear fragments (similarly to how a particle 
may lit up several detector cells). The required $n_\mathrm{tries}$
only depends on the number of local minima of $\Omega_R$ that reflects
the event's global structure (number, width of jets, etc.)
but not on $N_\mathrm{part}$.

The simplest OJF-based implementation of OJD with a fixed $n_\mathrm{tries}$
for all events is faster than KTCLUS for
$N_\mathrm{part} > 90\sqrt{n_\mathrm{tries}}$.
Note that the values above 7 for $n_\mathrm{tries}$ seem to be rarely
warranted, and for a substantial fraction of events very low values are
in fact sufficient. We have not explored this option, focusing instead on
a more significant optimization described below. 

\textbf{8.}\
It is important to appreciate that whereas any modification
of the $k_\mathrm{T}$ algorithm beyond an equivalent code transformation
would have to be treated as an entirely new jet definition,
OJD is formulated without reference to any specific implementation,
so once a reliable minimization algorithm is found, it can be used to control
the quality of other implementations designed for speed. 

Useful modifications result from allowing a misidentification of the global
minimum for a fraction of events, with the quality of the 
entire data processing procedure controlled via Fisher's information
$\left\langle {f_{{\mathrm{opt}}}^{\mathrm{2}} } \right\rangle $.
A simple such optimization can be implemented entirely using the routines
from the OJF library; it relies on the well-known fact that the jet structure
is often determined by the most energetic particles: Select the most
energetic particles (a skeleton event), and precluster them by running
the minimization routine. Then add the remaining particles with random
values of $z_{aj}$ and run the minimization again. With a threshold of 2 GeV
to select the energetic particles, $n_\mathrm{tries}=5$ at the
preclustering phase and $n_\mathrm{tries}$ =1 at the final stage,
only a 1\% change was observed for $\delta M_{{\mathrm{exp}}}$
(curiously, an improvement) whereas the speed much increased,
with the dependence of the time per event on $N_\mathrm{part}$ now given
roughly by
%
\begin{equation}\label{eq-11}
2.5 \times 10^{-2}  \times N_{\mathrm{part}}
\end{equation}
%
with a hint at a slower growth at large $N_\mathrm{part}$.
This is faster than KTCLUS starting from $N_\mathrm{part}\approx140$,
and the speed advantage increases sharply for higher $N_\mathrm{part}$:
for $N_\mathrm{part}\approx200$ this is twice as fast as
KTCLUS, and  an extrapolation to $N_\mathrm{part}\approx1000$
yields the factor of 50.

The dramatically better behavior of OJF at large $N_\mathrm{part}$ makes it
a candidate for work at the level of detector cells, perhaps even 
on-line (note that all $n_\mathrm{tries}$ minimization
attempts can be done in parallel).

The OJF library implements the first
minimization algorithm found to run acceptably fast.
Better algorithms may be found once the OJD/OJF is explored further.

\textbf{9.}\
To summarize, a conclusive method to compare jet algorithms
is based on evaluation of Fisher's information.
In the considered model measurement, OJD is equivalent to the $k_\mathrm{T}$
definition in physical quality, and an implementation of OJD 
is increasingly faster than KTCLUS at large
$N_\mathrm{part}$
starting from $N_\mathrm{part}\approx140$.
Moreover, OJD is defined in a theoretically 
preferred fashion and is supported by a systematic theory with new 
options for improvement of jets-based measurements.
All this positions OJD as a candidate for a standard jet definition
for the next generation of HEP experiments.

We thank A.\ Czarnecki and the two referees for useful criticisms.
FT thanks A.\ Czarnecki for hospitality at the University of 
Alberta (Canada) where a part of this work was done.
This work was supported in parts by the Natural Sciences and 
Engineering Research Council of Canada and the NATO grant PST.CLG.977751.


\begin{thebibliography}{99}
\bibitem{r1}
F.\ V.\ Tkachov, Int.\ J.\ Mod.\ Phys.\ \textbf{A17}, 2783 (2002).
\bibitem{r2}
See e.g.\ R.\ Barlow, Rep.\ Prog.\ Phys.\ \textbf{36}, 1067 (1993).
\bibitem{r3}	
F.\ Dydak, talk at The IX Int.\ Workshop on High Energy Physics
(September 1994, Zvenigorod, Russia).
\bibitem{r4}	
G.\ Sterman and S.\ Weinberg, Phys.\ Rev.\ Lett.\ \textbf{39}, 1436 (1977).
\bibitem{r5}
N.\ A.\ Sveshnikov and F.\ V.\ Tkachov, Phys.\ Lett.\ \textbf{B382},
403 (1996).
\bibitem{r6}
S.\ D.\ Ellis et al., in: Research Directions for the Decade, Snowmass 1990.
Singapore: World Scientific, 1992.
\bibitem{r7}
J.\ B.\ Babcock and R.\ E.\ Cutkosky, Nucl.\ Phys.\ \textbf{B176}, 113 (1980).
\bibitem{r8}
S.\ Brandt et al., Phys.\ Lett.\ \textbf{12}, 57 (1964);  
E.\ Farhi, Phys.\ Rev.\ Lett.\ \textbf{39}, 1587 (1977).
\bibitem{r9}
S.\ Moretti, L.\ Lonnblad and T.\ Sjostrand, JHEP \textbf{9808}, 1 (1998).
\bibitem{r10}
T.\ Sjostrand, Comp.\ Phys.\ Comm.\ \textbf{28}, 229 (1983).
\bibitem{r11}
JADE collaboration, Z.\ Phys.\ \textbf{C33}, 23 (1986).
\bibitem{r12}
S.\ Bethke et al., Nucl.\ Phys.\ \textbf{B370}, 310 (1992).
\bibitem{r13}
S.\ Catani et al., Phys.\ Lett.\ \textbf{269B}, 432 (1991).
\bibitem{r14}
S.\ Youssef, Comp.\ Phys.\ Comm.\ \textbf{45}, 423 (1987).
\bibitem{r15}
E.\ L.\ Berger et al., e-Print hep-ph/0201146.
\bibitem{r16}
F.\ V.\ Tkachov, Phys.\ Rev.\ Lett.\ \textbf{73}, 2405 (1994);
Erratum, \textbf{74}, 2618 (1995).
\bibitem{r17}
F.\ V.\ Tkachov, Int.\ J.\ Mod.\ Phys.\ \textbf{A12}, 5411 (1997).
\bibitem{r18}
F.\ V.\ Tkachov, Part.\ Nucl., Letters \textbf{2[111]}, 28 (2002).
\bibitem{r19}
N.\ Amos et al., contribution to CHEP95,
URL: http://www.hep.net/chep95/html/papers/p155/.
\bibitem{r20}
P.\ C.\ Bhat, H.\ Prosper, and S.\ S.\ Snyder, Int.\ J.\ Mod.\ Phys.\
\textbf{ A13}, 5113 (1998).
\bibitem{r21}
C.\ F.\ Berger et al., e-Print hep-ph/0202207.
\bibitem{r23}
D.\ Yu.\ Grigoriev and F.\ V.\ Tkachov, e-Print hep-ph/9912415; 
E.\ Jankowski, D.\ Yu.\ Grigoriev and F.\ V.\ Tkachov, to be submitted
to Comp.\ Phys.\ Comm.
\bibitem{r24}
The FORTRAN code ver.\ OJF\_014 is publicly available from
http://www.inr.ac.ru/$^\sim$ftkachov/projects/jets/.
\bibitem{r25}
http://www.oberon.ch.
\bibitem{r26}
The first realistic test was run by P.\ Achard (L3, CERN) in 1999 with
a sample of about $10^5$ events.
\bibitem{r27}
F.\ V.\ Tkachov, e-Print hep-ph/0111035.
\bibitem{r28}
Technical report CERN-EP-2000-099.
\bibitem{r29}
E.\ Jankowski and F.\ V.\ Tkachov, in preparation.
\bibitem{r30}
T.\ Sjostrand et al., Comp.\ Phys.\ Comm.\ \textbf{135}, 238 (2001).
\bibitem{r31}
http://hepwww.rl.ac.uk/theory/seymour/ktclus/
\bibitem{r32}
Run II Jet Physics, e-Print hep-ex/0005012v2, Sec.\ 4.3.2.
\bibitem{r33}
P.\ W.\ Bopp, Z.\ Phys.\ \textbf{C3}, 171 (1979).
\bibitem{r34}
W.\ Bartel at al.\ (JADE Collab.), Phys.\ Lett.\ \textbf{B91}, 142 (1980).
\end{thebibliography}
\end{document}